\documentclass[reprent,aps,prb,twocolumn,superscriptaddress,nolongbibliography]{revtex4-2}

\usepackage{amsmath,amssymb}
\usepackage{graphicx}
\usepackage{bm}
\usepackage{multirow}
\usepackage{hyperref}
\usepackage{wasysym}

\begin{document}

\title{Scalar spin chirality induced by a circularly polarized electric field\\ in a classical kagome magnet}
\author{Ryota Yambe}
\email{ryota_yambe@alumni.u-tokyo.ac.jp}
\affiliation{Department of Applied Physics, The University of Tokyo, Tokyo 113-8656, Japan }
\author{Satoru Hayami}
\email{hayami@phys.sci.hokudai.ac.jp}
\affiliation{Graduate School of Science, Hokkaido University, Sapporo 060-0810, Japan}

\begin{abstract}
Noncoplanar magnetic states with a scalar spin chirality have been intensively studied in condensed matter physics, since they exhibit fascinating physical phenomena.
We theoretically propose the generation of such noncoplanar magnetic states by using a circularly polarized electric field.
By performing the micromagnetic simulation, we investigate a time evolution of a classical kagome magnet irradiated by the circularly polarized electric field.
As a result, we find that the noncoplanar magnetic states are induced as a nonequilibrium steady state irrespective of the ground-state spin configurations.
We show that the induced scalar spin chirality is controlled by the amplitude, frequency, and polarization of the electric field.
In addition, we clarify that the mechanism of the noncoplanar magnetic
 states is accounted for by effective field-induced three-spin interactions by adopting the Floquet formalism in the high-frequency regime. 
We also show a condition to enhance the scalar spin chirality.
Our results present a new reference for controlling the noncoplanar magnetic states and related phenomena by the circularly polarized electric field.
\end{abstract}

\maketitle

\section{Introduction}
\label{sec:introduction}

Engineering of physical properties by a time-periodic field has attracted much attention in various fields of condensed matter physics.
A time evolution by the time-periodic field has been studied by the Floquet formalism, which provides a framework for understanding the time-periodic evolution by using a time-independent Floquet Hamiltonian~\cite{eckardt2017colloquium,oka2019floquet,rudner2020floquet}.
Since the Floquet Hamiltonian exhibits a variety of effective terms depending on the amplitude, frequency, and polarization of the time-periodic field, its radiation becomes a source of rich physical properties compared to the radiation of a static field.

In magnetic systems, the effect of the time-periodic field radiation appears as field-induced magnetic interactions in the Floquet Hamiltonian.  
For example, the radiation of the time-periodic electric field on Mott insulators induces and modifies exchange interactions~\cite{mentink2015ultrafast, PhysRevLett.121.107201, PhysRevB.99.205111,bukov2016schrieffer,chaudhary2019orbital, PhysRevLett.126.177201, PhysRevB.103.L100408, PhysRevB.104.214413, PhysRevB.105.085144,kumar2022floquet, PhysRevB.105.L180414, PhysRevResearch.4.L032036} and multiple-spin interactions~\cite{PhysRevB.96.014406,claassen2017dynamical,quito2021polarization, PhysRevB.105.054423}.
Such changes in the interactions also occur in spin systems~\cite{PhysRevB.90.214413, PhysRevB.90.085150,sato2014floquet, PhysRevLett.117.147202, PhysRevLett.128.037201,higashikawa2018floquet, PhysRevB.108.064420} and other itinerant electron systems~\cite{PhysRevLett.119.147202} as a result of the  time-periodic field radiation.
It was theoretically shown that physical properties such as a magnetization~\cite{PhysRevB.90.214413,PhysRevB.90.085150,higashikawa2018floquet} and states of the matter such as a quantum spin liquid~\cite{PhysRevResearch.4.L032036,claassen2017dynamical,PhysRevB.105.054423,sato2014floquet}, a helical state~\cite{PhysRevLett.117.147202,higashikawa2018floquet}, and a skyrmion~\cite{PhysRevLett.128.037201,PhysRevLett.119.147202} can be controlled by the field-induced magnetic interactions.
This indicates that the time-periodic field radiation is one of the powerful tools to control magnetic states and related electromagnetic responses.

In this paper, we propose the further intriguing possibility of controlling magnetic properties by a time-periodic field. We especially focus on a field-induced scalar spin chirality under noncoplanar magnetic states, which becomes the origin of the topological Hall/Nernst effect~\cite{Ohgushi_PhysRevB.62.R6065,taguchi2001spin, Neubauer_PhysRevLett.102.186602, Shiomi_PhysRevB.88.064409,kurumaji2019skyrmion, Hirschberger_PhysRevLett.125.076602}.
For that purpose, we radiate a circularly polarized electric field on a collinear or coplanar magnet without the scalar spin chirality in the kagome-lattice structure, where we suppose that the electric field is coupled to spins via a spin-dependent electric polarization mechanism.
By calculating a time evolution based on the micromagnetic simulation, we show that the circularly polarized electric field induces noncoplanar magnetic states and their scalar spin chirality is controlled by the amplitude, frequency, and polarization of the electric field.  
In addition, we derive a time-independent Floquet Hamiltonian based on the Floquet formalism in order to investigate the origin of the field-induced scalar spin chirality.
As a result, we clarify that the scalar spin chirality originates from effective field-induced three-spin interactions under the circularly polarized electric field.
We also discuss a way of enhancing the scalar spin chirality via the time-periodic
 electric field.

The paper is organized as follows.
In Sec.~\ref{sec:static_model}, we introduce a static model for a classical kagome magnet and discuss the ground states under no external fields.
Then, we introduce a dynamical Hamiltonian by taking into account the effect of a circularly polarized electric field in Sec.~\ref{sec:dynamical_model_hamiltonian}.
We also outline methods to analyze the dynamical model based on the micromagnetic simulation and the Floquet formalism in Secs.~\ref{sec:dynamical_model_LLG} and \ref{sec:dynamical_model_Floquet}, respectively.
In Sec.~\ref{sec:simulation}, we show the origin of the field-induced scalar spin chirality and its dependence on model parameters by performing the micromagnetic simulations. 
Section~\ref{sec:summary} summarizes the paper and discusses a possible experimental situation.

\section{Static model}
\label{sec:static_model}

In Sec.~\ref{sec:static_model_hamiltonian}, we introduce a static Hamiltonian for a classical kagome magnet.
The ground states of the static model are shown in Sec.~\ref{sec:static_model_GS}.

\subsection{Hamiltonian}
\label{sec:static_model_hamiltonian}

\begin{figure}[t!]
\begin{center}
\includegraphics[width=1.0\hsize]{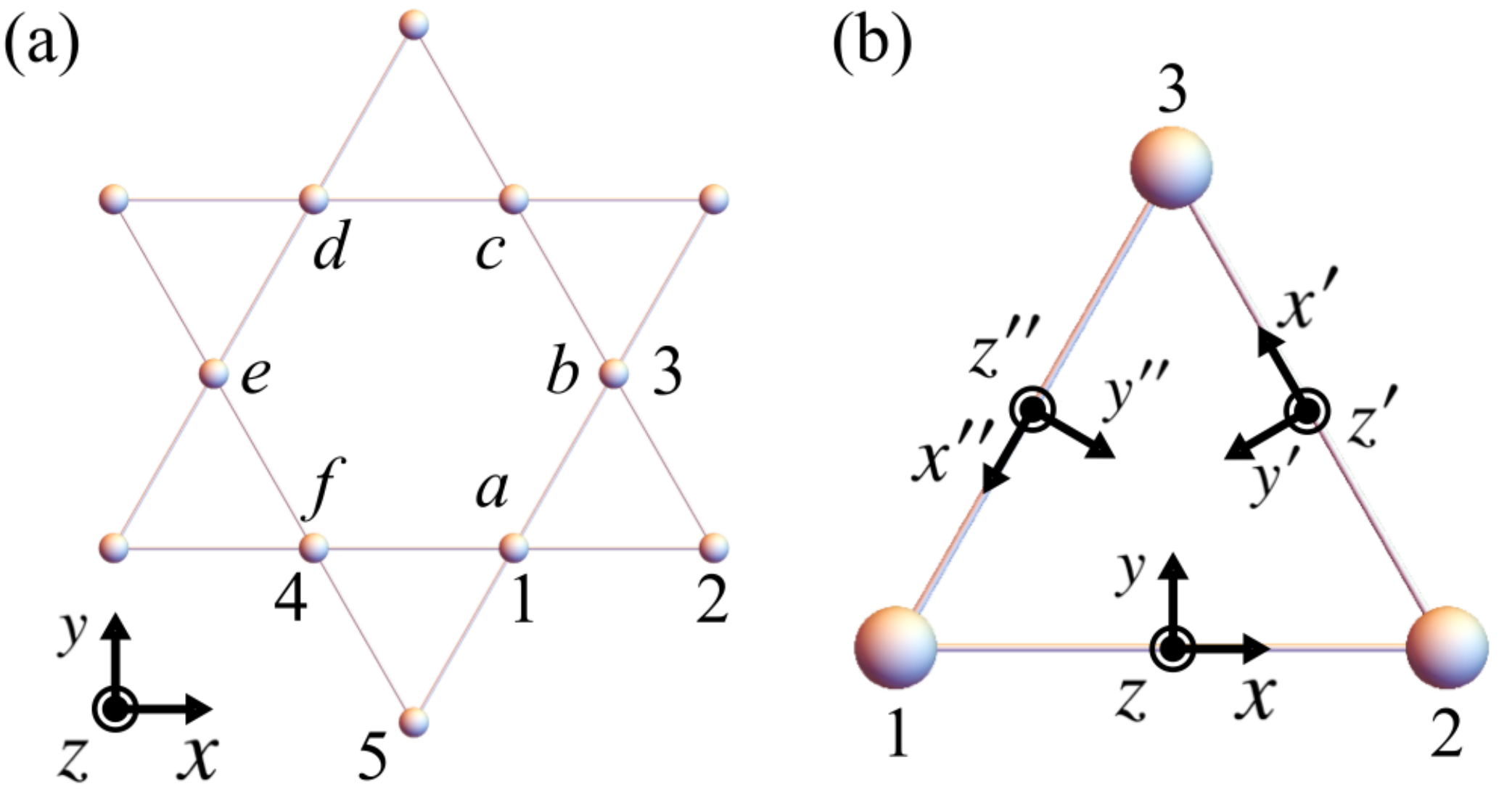} 
\caption{\label{fig:kagome}
(a) Kagome network consisting of three sublattices 1--3 under the point group $\bar{6}m2$. 
The indices 1--5 and $a$--$f$ represent the site indices used to denote the interaction and the electric polarization; see the main text in detail.
(b) Local cartesian spin coordinates $(x,y,z)$ for the $\langle 1,2 \rangle$ bond, $(x',y',z')$ for the $\langle 2,3 \rangle$ bond, and $(x'',y'',z'')$ for the $\langle 3,1 \rangle$ bond~\cite{PhysRevB.108.064420}.
}
\end{center}
\end{figure}

We consider a static model for a classical kagome magnet with the point group $\bar{6}m2$ by implicitly supposing the different sizes of the upward and downward triangles, which is the so-called breathing kagome magnet.
The Hamiltonian is given by
\begin{align}
\label{eq:static_hamiltonian}
\mathcal{H}_0&=\sum_{\bigtriangleup}\sum_{\alpha,\beta}J^{\alpha\beta}_{\bigtriangleup}\left( 
m^\alpha_1m^\beta_2
+m^{\alpha'}_2m^{\beta'}_3
+m^{\alpha''}_3m^{\beta''}_1
\right) \nonumber \\
&+\sum_{\bigtriangledown}\sum_{\alpha,\beta}J^{\alpha\beta}_{\bigtriangledown}\left( 
m^\alpha_1m^\beta_4
+m^{\alpha'}_4m^{\beta'}_5
+m^{\alpha''}_5m^{\beta''}_1
\right),
\end{align}
where $\boldsymbol{m}_j$ is the local magnetic moment at site $j$ with $|\boldsymbol{m}_j|=1$, $\alpha,\beta=x,y,z$, the summation $\sum_{\bigtriangleup (\bigtriangledown)}$ is taken over all the upward (downward) triangles on the kagome network, and sites 1, 2 and 3 (1, 4, and 5) on each upward (downward) triangle are labeled in the counterclockwise order [see Fig.~\ref{fig:kagome}(a)].
We use local cartesian spin coordinates $(x,y,z)$ for the $\langle 1,2 \rangle$ and $\langle 1,4 \rangle$ bonds, $(x',y',z')$ for the $\langle 2,3 \rangle$ and $\langle 4,5 \rangle$ bonds, and $(x'',y'',z'')$ for the $\langle 3,1 \rangle$ and $\langle 5,1 \rangle$ bonds, as shown in Fig.~\ref{fig:kagome}(b).
The nearest-neighbor interaction matrices for the upward and downward triangles are generally given by
\begin{align}
\label{eq:static_interaction_up}
J_{\bigtriangleup} &= \begin{pmatrix}
F^x_{\bigtriangleup} & D^z_{\bigtriangleup} & 0 \\
-D^z_{\bigtriangleup} & F^y_{\bigtriangleup} & 0 \\
0 & 0 & F^z_{\bigtriangleup}
\end{pmatrix},\\
\label{eq:static_interaction_down}
 J_{\bigtriangledown} &= \begin{pmatrix}
F^x_{\bigtriangledown} & D^z_{\bigtriangledown} & 0 \\
-D^z_{\bigtriangledown} & F^y_{\bigtriangledown} & 0 \\
0 & 0 & F^z_{\bigtriangledown}
\end{pmatrix},
\end{align}
respectively.
Here, $F^x_{\bigtriangleup(\bigtriangledown)}$, $F^y_{\bigtriangleup(\bigtriangledown)}$, and $F^z_{\bigtriangleup(\bigtriangledown)}$ are symmetric anisotropic exchange interactions, and $D^z_{\bigtriangleup(\bigtriangledown)}$ is the Dzyaloshinskii-Moriya (DM) interaction~\cite{dzyaloshinsky1958thermodynamic,moriya1960anisotropic} for upward (downward) triangles. 
The components in Eqs.~(\ref{eq:static_interaction_up}) and (\ref{eq:static_interaction_down}) satisfy the point group symmetry $\bar{6}m2$ of the lattice.

\subsection{Ground states}
\label{sec:static_model_GS}

\begin{figure}[t!]
\begin{center}
\includegraphics[width=1.0\hsize]{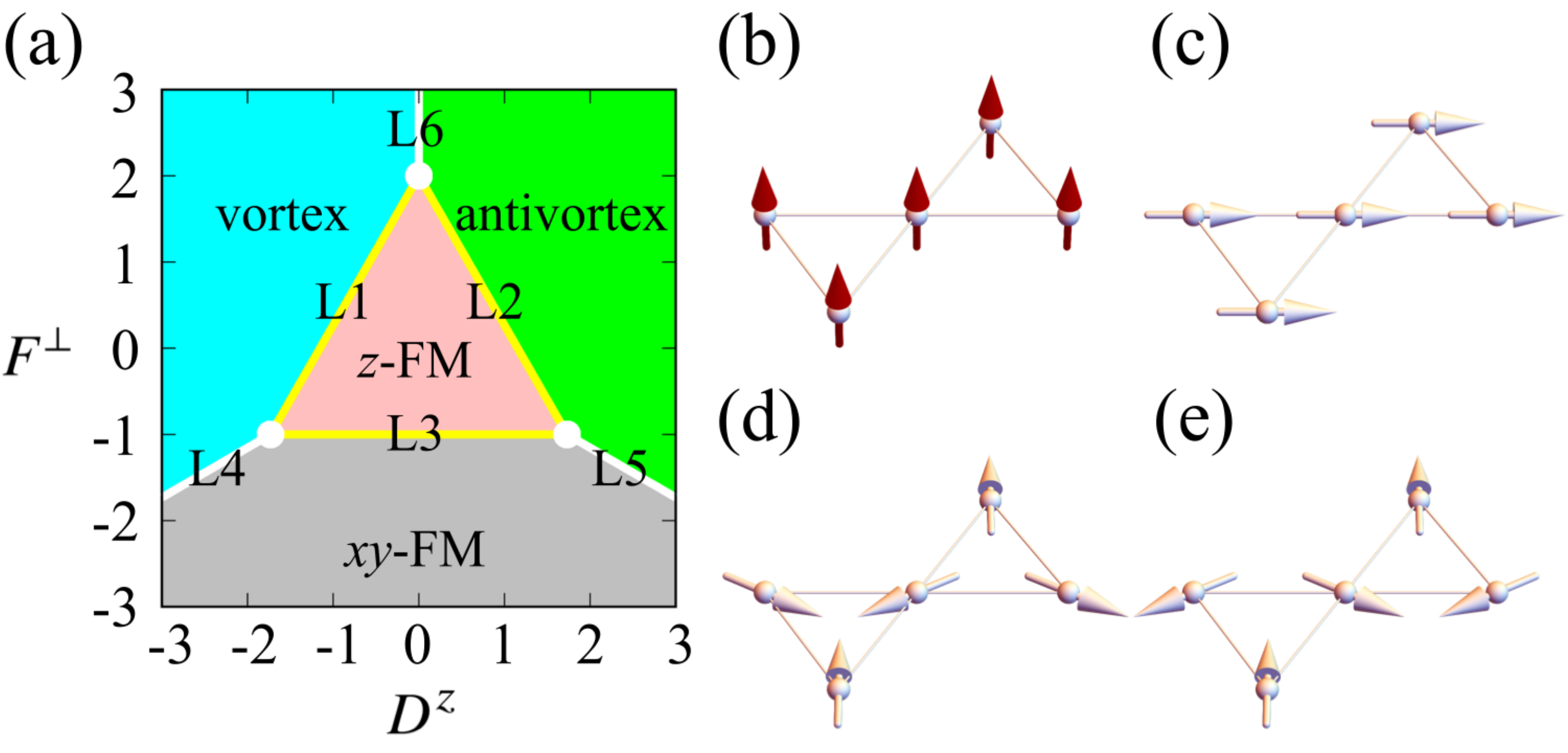} 
\caption{\label{fig:GS}
(a) Ground-state phase diagram for the static model in Eq.~(\ref{eq:static_hamiltonian}) with $F^z=-1$~\cite{PhysRevB.96.205126}: $z$-FM (pink region), $xy$-FM (gray region), vortex (cyan region), and antivortex (green region).
The phase boundaries L1-L6 are denoted by yellow and white lines.
Magnetic configurations for (b) $z$-FM, (c) $xy$-FM, (d) vortex, and (e) antivortex phases.
The arrows and their colors show the local magnetic moments and the $z$ components, respectively.
The red and white colors represent the local magnetic moments with $m^z_j=1$ and $m^z_j=0$, respectively.
}
\end{center}
\end{figure}

We discuss the ground states of the static model in Eq.~(\ref{eq:static_hamiltonian}).
We simplify the static model by setting $F^z_{\bigtriangleup}=F^z_{\bigtriangledown}\equiv F^z$,  $F^x_{\bigtriangleup}=F^x_{\bigtriangledown}=F^y_{\bigtriangleup}=F^y_{\bigtriangledown}\equiv F^\perp$, and $D^z_{\bigtriangleup}=D^z_{\bigtriangledown}\equiv D^z$; this model corresponds to conventional kagome model with the point group $6/mmm$; the effect of the difference between the upward and downward triangles will be considered in a dynamical model in Sec.~\ref{sec:dynamical_model_hamiltonian}.
In addition, it is noted that the model has accidental global U(1) symmetry around the $z$ axis in spin space.
Then, the ground states are obtained by an analytical calculation~\cite{PhysRevB.96.205126}.

We show a ground-state phase diagram on the $D^z$-$F^{\perp}$ plane for the static model with $F^z=-1$ in Fig.~\ref{fig:GS}(a).
There are four long-range ordered phases with an ordering wave vector of $\bm{q}=\bm{0}$, whose three-sublattice magnetic configurations are described by using the magnetic configurations on an upward triangle, $\bm{m}=(\bm{m}_1,\bm{m}_2,\bm{m}_3)$, as follows:
\begin{itemize}
\item[(i)] Out-of-plane ferromagnetic ($z$-FM) phase.
The magnetic configuration is given by
\begin{align}
\label{eq:conf_z}
\bm{m}^{ z\text{-FM} } &=\pm(0,0,1,0,0,1,0,0,1),
\end{align}
whose ground-state energy per unit cell is given by
\begin{align}
\label{eq:E_z}
E^{z\text{-FM}} &=6F^z.
\end{align}
There is a degeneracy between the states with positive and negative $z$ components.
The $z$-FM configuration with positive $z$ components is shown in Fig.~\ref{fig:GS}(b).
\item[(ii)] In-plane ferromagnetic ($xy$-FM) phase.
The magnetic configuration is given by
\begin{align}
\label{eq:conf_xy}
\bm{m}^{ xy\text{-FM} } =&(\cos\theta,\sin\theta,0,\cos\theta,\sin\theta,0,\nonumber\\
&\cos\theta,\sin\theta,0),
\end{align}
with an angle $\theta$ $(0<\theta\le2\pi)$.
The ground-state energy per unit cell is given by
\begin{align}
\label{eq:E_xy}
E^{xy\text{-FM}} &=6F^\perp,
\end{align}
which is independent of $\theta$ due to the U(1) symmetry of the model.
The $xy$-FM configuration for $\theta=0$ is shown in Fig.~\ref{fig:GS}(c).
\item[(iii)] Vortex (Antivortex) phase.
The magnetic configuration for the vortex (antivortex) state is given by
\begin{align}
\label{eq:conf_vortex}
\bm{m}^{v(av)} =&
\{ \sin(\theta+\sigma2\pi/3),\cos(\theta+\sigma2\pi/3),0,\nonumber\\
&\sin(\theta-\sigma2\pi/3),\cos(\theta-\sigma2\pi/3),0, \nonumber\\
&\sin\theta,\cos\theta,0\},
\end{align}
with $\sigma=+1$ ($-1$) and $\theta$ $(0<\theta\le2\pi)$.
The ground-state energy per unit cell is given by
\begin{align}
\label{eq:E_vortex}
E^{v} &=3(-F^\perp+\sigma \sqrt{3}D^z).
\end{align}
The negative (positive) $D^z$ favors the vortex (antivortex) state.
Similar to the $xy$-FM state,  the ground-state energy is independent of $\theta$ due to the U(1) symmetry of the model.
The vortex and antivortex configurations for $\theta=0$ are shown in Figs.~\ref{fig:GS}(d) and \ref{fig:GS}(e), respectively.
\end{itemize}
Among the four states, the $z$-FM and $xy$-FM states are categorized into the collinear state, while the vortex and antivortex states are categorized into the coplanar state.

The ground states at the phase boundaries are given by superposing the $z$-FM, $xy$-FM, vortex, and antivortex states.
At the phase boundary L1, since the energies of the $z$-FM and vortex states are degenerate, the ground state is given by superposing them as
\begin{align}
\label{eq:conf_L1}
\bm{m}^{ {\rm L1}} &=c_{z}\bm{m}^{ z\text{-FM}}+c_{v}\bm{m}^{v},
\end{align}
with $c_{z}^2+c_{v}^2=1$.  
The magnetic configurations for $c_{z}\neq0$ and $c_{v}\neq0$ are categorized into the noncoplanar state with nonzero scalar spin chirality, i.e., $\bm{m}_1 \cdot (\bm{m}_2 \times \bm{m}_3) \neq 0$ and $\bm{m}_1 \cdot (\bm{m}_4 \times \bm{m}_5) \neq 0$, which we call a scalar chiral state.
Similar to the phase boundary L1, the ground state at the phase boundary L2 becomes a noncoplanar magnetic state by superposing the $z$-FM and antivortex states. 
At the phase boundary L3, a superposition of the $z$-FM and $xy$-FM states becomes the ground state, which is categorized into the collinear state.  
At the phase boundaries L4, L5, and L6, magnetic states with ordering wave vectors $\bm{q}\neq\bm{0}$ become the ground states~\cite{PhysRevB.96.205126}.

\section{Dynamical model}
\label{sec:dynamical_model}

We introduce a dynamical Hamiltonian for the classical kagome magnet irradiated by a circularly polarized electric field in Sec.~\ref{sec:dynamical_model_hamiltonian}.
Then, we outline two methods to analyze the dynamical model.
One is the analysis based on the Landau-Lifshitz-Gilbert (LLG) equation in Sec.~\ref{sec:dynamical_model_LLG} and the other is the analysis based on the Floquet formalism in Sec.~\ref{sec:dynamical_model_Floquet}.

\subsection{Hamiltonian}
\label{sec:dynamical_model_hamiltonian}

We consider the effect of the circularly polarized electric field on the classical kagome magnet.
The Hamiltonian under the circularly polarized electric field is given by 
\begin{align}
\label{eq:dynamical_hamiltonian}
\mathcal{H}(t) = \mathcal{H}_0 -\bm{E}(t)\cdot\bm{P},
\end{align}
where $\mathcal{H}_0$ is given in Eq.~(\ref{eq:static_hamiltonian}) and $\bm{P}$ represents an electric polarization coupled to the circularly polarized electric field $\bm{E}(t) = E_0(\delta\cos\Omega t,-\sin\Omega t ,0)$; $E_0$ is the amplitude of the field, $\Omega$ is the frequency, and $\delta=+1$ $(-1)$ represents a right circular polarization (RCP) [left circular polarization (LCP)]. 
We suppose that the electric polarization originates from spin-dependent electric dipoles on the nearest-neighbor bonds as
\begin{align}
\bm{P} = \sum_{\bigtriangleup}(\bm{p}_{12}+\bm{p}_{23}+\bm{p}_{31})+\sum_{\bigtriangledown}(\bm{p}_{14}+\bm{p}_{45}+\bm{p}_{51}),
\end{align}
with
\begin{align}
\bm{p}_{12} &= -\lambda_{\bigtriangleup} \bm{e}_{12} \times (\bm{m}_1\times \bm{m}_2),\\
\bm{p}_{23} &= -\lambda_{\bigtriangleup} \bm{e}_{23} \times (\bm{m}_2\times \bm{m}_3),\\
\bm{p}_{31} &= -\lambda_{\bigtriangleup} \bm{e}_{31} \times (\bm{m}_3\times \bm{m}_1),\\
\bm{p}_{14} &= -\lambda_{\bigtriangledown} \bm{e}_{14} \times (\bm{m}_1\times \bm{m}_4),\\
\bm{p}_{45} &= -\lambda_{\bigtriangledown} \bm{e}_{45} \times (\bm{m}_4\times \bm{m}_5),\\
\bm{p}_{51} &= -\lambda_{\bigtriangledown} \bm{e}_{51} \times (\bm{m}_5\times \bm{m}_1).
\end{align}
Here, the electric dipole $\bm{p}_{jk}$ for the nearest-neighbor bond $\langle j,k \rangle$ is induced by the spin-current mechanism~\cite{Katsura_PhysRevLett.95.057205,Mostovoy_PhysRevLett.96.067601,SergienkoPhysRevB.73.094434}; $\lambda_{\bigtriangleup (\bigtriangledown)}$ is the magnetoelectric coupling constant and $\bm{e}_{jk}$ is the unit vector from the site $j$ to the site $k$.
It is noted that the form of $\bm{p}_{12}$ ($\bm{p}_{14}$) satisfies the point group symmetry $mm2$ of the bond  $\langle 1,2 \rangle$ ($\langle 1,4 \rangle$)~\cite{PhysRevB.83.174432}, and the forms of $\bm{p}_{23}$ and $\bm{p}_{31}$ ($\bm{p}_{45}$ and $\bm{p}_{51}$) are determined to satisfy the threefold rotational symmetry of the upward (downward) triangle under the $\bar{6}m2$ symmetry.
It is noted that there is no symmetry constraint between $\lambda_{\bigtriangleup}$ and $\lambda_{\bigtriangledown}$ under the point group $\bar{6}m2$ (breathing kagome structure), while $\lambda_{\bigtriangleup}$ is equivalent to $\lambda_{\bigtriangledown}$ under the point group $6/mmm$ (conventional kagome structure).

\subsection{Landau-Lifshitz-Gilbert equation}
\label{sec:dynamical_model_LLG}

We calculate a time evolution of the dynamical Hamiltonian $\mathcal{H}(t)$ in Eq.~(\ref{eq:dynamical_hamiltonian}) by numerically solving the LLG equation.
The LLG equation is given by
\begin{align}
\label{eq:LLG}
\frac{d\bm{m}_j}{dt}=&-\frac{\gamma}{1+\alpha_{\rm G}^2}\left[ \bm{m}_j\times\bm{B}^{\rm eff}_j(t) \right. \nonumber \\
& \left.+ \alpha_{\rm G}\bm{m}_j\times\{\bm{m}_j\times\bm{B}^{\rm eff}_j(t)\} \right],
\end{align}  
with the gyromagnetic ratio $\gamma$, the Gilbert damping constant $\alpha_{\rm G}$, and the effective magnetic field $\bm{B}^{\rm eff}_{j}(t)=-\partial\mathcal{H}(t)/\partial \bm{m}_{j}$.
The first term represents the precession around the effective magnetic field, and the second term  describes the relaxation to the effective magnetic field.

The effective magnetic field for $\bm{m}_j$ is given by 
\begin{align}
\label{eq:Beff}
\bm{B}^{\rm eff}_j (t) = &-\frac{\partial \mathcal{H}_0}{\partial \bm{m}_{j}
}
-\lambda_{\bigtriangleup}\sum_{k_{\rm NN}\in\bigtriangleup}  \bm{m}_{k_{\rm NN}} \times \{\bm{E}(t) \times \bm{e}_{jk_{\rm NN}}\} \nonumber \\
&-\lambda_{\bigtriangledown}\sum_{k_{\rm NN}\in\bigtriangledown}  \bm{m}_{k_{\rm NN}} \times \{\bm{E}(t) \times \bm{e}_{jk_{\rm NN}}\},
\end{align} 
where $k_{\rm NN}\in\bigtriangleup(\bigtriangledown)$ represents the nearest-neighbor site on the upward (downward) triangle.
The first term is independent of time, while the second and third terms are time-periodic with a period of $T=2\pi/\Omega$.

To judge whether noncoplanar states are induced by the electric field, we calculate the scalar spin chirality at each time as
\begin{align}
\label{eq:sc_t}
\chi_{\rm sc}(t) &=  \frac{1}{N}\sum_{\bigtriangleup}\boldsymbol{m}_1(t)\cdot \left\{\boldsymbol{m}_2(t)\times\boldsymbol{m}_3(t)\right\} \nonumber \\
& 
+ \frac{1}{N}\sum_{\bigtriangledown}\boldsymbol{m}_1(t)\cdot \left\{ \boldsymbol{m}_4(t)\times\boldsymbol{m}_5(t) \right\},
\end{align}
where $N$ is the number of unit cells.
After a long time $t_0$, the system reaches a non-equilibrium steady state (NESS).
To characterize the NESS, we calculate an averaged scalar spin chirality, which is given by
\begin{align}
\label{eq:sc}
\chi_{\rm sc} &= \frac{1}{N_{\rm ave}}\sum_{n=1}^{N_{\rm ave}} \chi_{\rm sc}(t_0+n\Delta).
\end{align} 
Here, $N_{\rm ave}$ and $\Delta$ represent the number of samples and the time step, respectively.

\subsection{Floquet analysis}
\label{sec:dynamical_model_Floquet}

Since the LLG equation in Eq.~(\ref{eq:LLG}) is time-periodic, we can adopt the Floquet formalism for the classical systems~\cite{higashikawa2018floquet}, which results in a time-independent effective magnetic field and Floquet Hamiltonian.
By solving the LLG equation within the Floquet formalism, one obtains a magnetic state with a local energy minimum of the Floquet Hamiltonian.
The obtained magnetic state is expected to be consistent with the NESS when the frequency $\Omega$ is large and the amplitude $E_0$ is small compared to the energy scale of the static model.

With this in mind, we perform the Floquet analysis when the frequency $\Omega$ is large enough compared to the energy scale of the static model in Eq.~(\ref{eq:static_hamiltonian}).
By using the high-frequency expansion in the Floquet formalism, a time-independent effective magnetic field up to $\Omega^{-1}$ is obtained as
\begin{align} 
\label{eq:Beff_Floquet}
\tilde{\bm{B}}^{\rm eff}_j &= \bm{B}^{\rm eff}_{j,0} + \sum_{n > 0}\frac{i\gamma [\bm{B}^{\rm eff}_{j,-n},\bm{B}^{\rm eff}_{j,+n}]}{n \Omega (1+\alpha_{\rm G}^2)}.
\end{align} 
Here, $n$ is an integer and $\bm{B}^{\rm eff}_{j,n} = T^{-1}\int^{T}_{0} dt\bm{B}^{\rm eff}_{j}(t) e^{in \Omega t}$ is the Fourier component of the time-periodic effective magnetic field.
The relation $[A , B ]$ is defined as
\begin{align} 
\label{eq:commun}
&[\bm{B}^{\rm eff}_{j,-n},\bm{B}^{\rm eff}_{j,+n}] \nonumber \\
&= \bm{B}^{\rm eff}_{j,-n}\times\bm{B}^{\rm eff}_{j,+n} +\sum_{k\neq j} \left[ ( \bm{B}^{\rm eff}_{k,-n} \cdot \bm{L}_k ) \bm{B}^{\rm eff}_{j,+n} \right.\nonumber \\
&\left.- ( \bm{B}^{\rm eff}_{k,+n} \cdot \bm{L}_k ) \bm{B}^{\rm eff}_{j,-n} \right] + \mathcal{O}(\alpha_{\rm G}),
\end{align} 
with $L_k^\alpha = - \sum_{\beta,\eta} \epsilon_{\alpha\beta\eta} m_{k}^\beta (\partial / \partial m_{k}^\eta) $ $(\alpha,\beta,\eta = x,y,z)$.
We ignore terms proportional to $\alpha_{\rm G}$ by assuming small $\alpha_{\rm G}$.

The Fourier components of $\bm{B}^{\rm eff}_j (t)$ in Eq.~(\ref{eq:Beff}) are given by
\begin{align}
\label{eq:B0}
\bm{B}^{\rm eff}_{j,0} =&-\frac{\partial \mathcal{H}_0}{\partial \bm{m}_j},\\
\label{eq:B1}
\bm{B}^{\rm eff}_{j,\pm1} =& 
-\lambda_{\bigtriangleup}\sum_{k_{\rm NN}\in\bigtriangleup}  \bm{m}_{k_{\rm NN}} \times (\tilde{\bm{E}}_{\pm 1} \times \bm{e}_{jk_{\rm NN}}) \nonumber \\
&-\lambda_{\bigtriangledown}\sum_{k_{\rm NN}\in\bigtriangledown}  \bm{m}_{k_{\rm NN}} \times (\tilde{\bm{E}}_{\pm 1} \times \bm{e}_{jk_{\rm NN}}),
\end{align}
with $\tilde{\bm{E}}_{\pm 1}=E_0(\delta,\pm i,0)/2$; the other Fourier components are zero. 
By substituting Eqs.~(\ref{eq:B0}) and (\ref{eq:B1}) into Eq.~(\ref{eq:Beff_Floquet}),  we obtain a time-independent effective magnetic field of the present model.

A time-independent Floquet Hamiltonian $\mathcal{H}^{\rm F}$ is constructed from a relation $\tilde{\bm{B}}^{\rm eff}_j=-\partial\mathcal{H}^{\rm F}/\partial \bm{m}_j$, which is given by 
\begin{align}
\label{eq:floquet_hamiltonian}
\mathcal{H}^{\rm F} =& \mathcal{H}_0 + \mathcal{T}_{\bigtriangleup}\sum_{\bigtriangleup}\bm{m}_1\cdot\bm{m}_2\times\bm{m}_3 + \mathcal{T}_{\bigtriangledown}\sum_{\bigtriangledown}\bm{m}_1\cdot\bm{m}_4\times\bm{m}_5 \nonumber\\
&+\mathcal{T}_{\hexagon}\sum_{\hexagon} [ 
S^z_a(\bm{m}_b\times\bm{m}_f)^z + S^z_b(\bm{m}_c\times\bm{m}_a)^z  \nonumber\\
& +S^z_c(\bm{m}_d\times\bm{m}_b)^z +S^z_d(\bm{m}_e\times\bm{m}_c)^z \nonumber \\
& +S^z_e(\bm{m}_f\times\bm{m}_d)^z + S^z_f(\bm{m}_a\times\bm{m}_e)^z],
\end{align}
with 
\begin{align}
\label{eq:T1}
 \mathcal{T}_{\bigtriangleup} &=-\frac{\sqrt{3}\gamma\delta(\lambda_{\bigtriangleup} E_0)^2}{4\Omega(1+\alpha_\mathrm{G})^2}, \\
 \label{eq:T2}
 \mathcal{T}_{\bigtriangledown} &=-\frac{\sqrt{3}\gamma\delta(\lambda_{\bigtriangledown} E_0)^2}{4\Omega(1+\alpha_\mathrm{G})^2}, \\
 \label{eq:T3}
 \mathcal{T}_{\hexagon}&=-\frac{\sqrt{3}\gamma\delta\lambda_{\bigtriangleup}\lambda_{\bigtriangledown} E_0^2}{4\Omega(1+\alpha_\mathrm{G})^2}.
\end{align}
The summation $\sum_{\hexagon}$ is taken over all the hexagons on the kagome network, and sites $a$, $b$, $c$, $d$, $e$, and $f$ on each hexagon are labeled in the counterclockwise order [see Fig.~\ref{fig:kagome}(a)].
The Floquet Hamiltonian includes the field-induced three-spin interactions with $\mathcal{T}_{\bigtriangleup}$, $\mathcal{T}_{\bigtriangledown}$, and $\mathcal{T}_{\hexagon}$ in addition to the static Hamiltonian $\mathcal{H}_0$.
In particular, the three-spin interactions with $\mathcal{T}_{\bigtriangleup}$ and $\mathcal{T}_{\bigtriangledown}$ are directly coupled to the scalar spin chirality.
The amplitude of the field-induced terms is controlled by the amplitude of the electric field $E_0$ and the frequency $\Omega$, and their sign is controlled by the polarization $\delta$.
It is noted that they cannot be induced by a linearly polarized electric field with $\delta=0$. 
Similar three-spin interactions were obtained in Mott insulators~\cite{PhysRevB.96.014406,claassen2017dynamical,quito2021polarization,PhysRevB.105.054423} and quantum spin systems~\cite{sato2014floquet,PhysRevB.108.064420}.

In the following calculations, we set $E_{\rm d}\equiv E_0\lambda_{\bigtriangleup}=-E_0\lambda_{\bigtriangledown}$ for simplicity, where the opposite sign for the upward and downward triangles is allowed by the $\bar{6}m2$ symmetry.
When the model is reduced to the conventional kagome model under the $6/mmm$ symmetry, $\lambda_{\bigtriangleup}$ is equivalent to $\lambda_{\bigtriangledown}$.
This situation does not induce the scalar spin chirality; see the results in Sec.~\ref{sec:simulation} for details.

\section{Simulation results}
\label{sec:simulation}

In this section, we show that the circularly polarized electric field induces the scalar spin chirality in NESSs by numerically solving the LLG equation. 
We consider the static model in Eq.~(\ref{eq:static_hamiltonian}) with a system size of $N=3^2$ under the periodic boundary condition~\footnote{We confirm that similar results are obtained in the model with a larger system size, e.g., $N=9^2$.}.
The energy scale and time scale are set as $E_{\rm GS}$ and $E_{\rm GS}^{-1}$, respectively; $E_{\rm GS}$ is the absolute value of the ground-state energy.  
Since we take $E_0\lambda_{\bigtriangleup}=-E_0\lambda_{\bigtriangledown}$, the relation as $\mathcal{T}\equiv\mathcal{T}_{\bigtriangleup}=\mathcal{T}_{\bigtriangledown}=-\mathcal{T}_{\hexagon}$ holds in the Floquet Hamiltonian in Eq.~(\ref{eq:floquet_hamiltonian}).
From the viewpoint of the effective interactions, one finds that the scalar spin chirality is not induced in the dynamical model for the conventional kagome magnet with $\lambda_{\bigtriangleup}=\lambda_{\bigtriangledown}$, since the effects of $\mathcal{T}_{\bigtriangleup}$, $\mathcal{T}_{\bigtriangledown}$ and $\mathcal{T}_{\hexagon}$ are canceled out. 
Thus, the difference between the upward and downward triangles is essential to induce the scalar spin chirality in kagome magnets.
In the LLG equation in Eq.~(\ref{eq:LLG}) we set $\gamma=1$ and $\alpha_{\rm G}=0.05$.
We calculate a time evolution for a long time, $t_0=80000E_{\rm GS}^{-1}$-$160000E_{\rm GS}^{-1}$, to obtain  NESSs.
After that, we calculate the averaged scalar spin chirality in Eq.~(\ref{eq:sc}) by setting $N_{\rm ave}=50000$ and $\Delta=0.002E_{\rm GS}^{-1}$.
We use an open software DifferentialEquations.jl~\cite{rackauckas2017differentialequations} to solve the LLG equation.

In the following, we discuss the effect of the circularly polarized electric field on the vortex and $z$-FM states in Secs.~\ref{sec:simulation_vortex} and \ref{sec:simulation_FM}, respectively. 
In Sec.~\ref{sec:simulation_parameter}, we show the static-parameter dependence of the field-induced scalar spin chirality.

\subsection{Radiation on a vortex state}
\label{sec:simulation_vortex}

\begin{figure}[t!]
\begin{center}
\includegraphics[width=1.0\hsize]{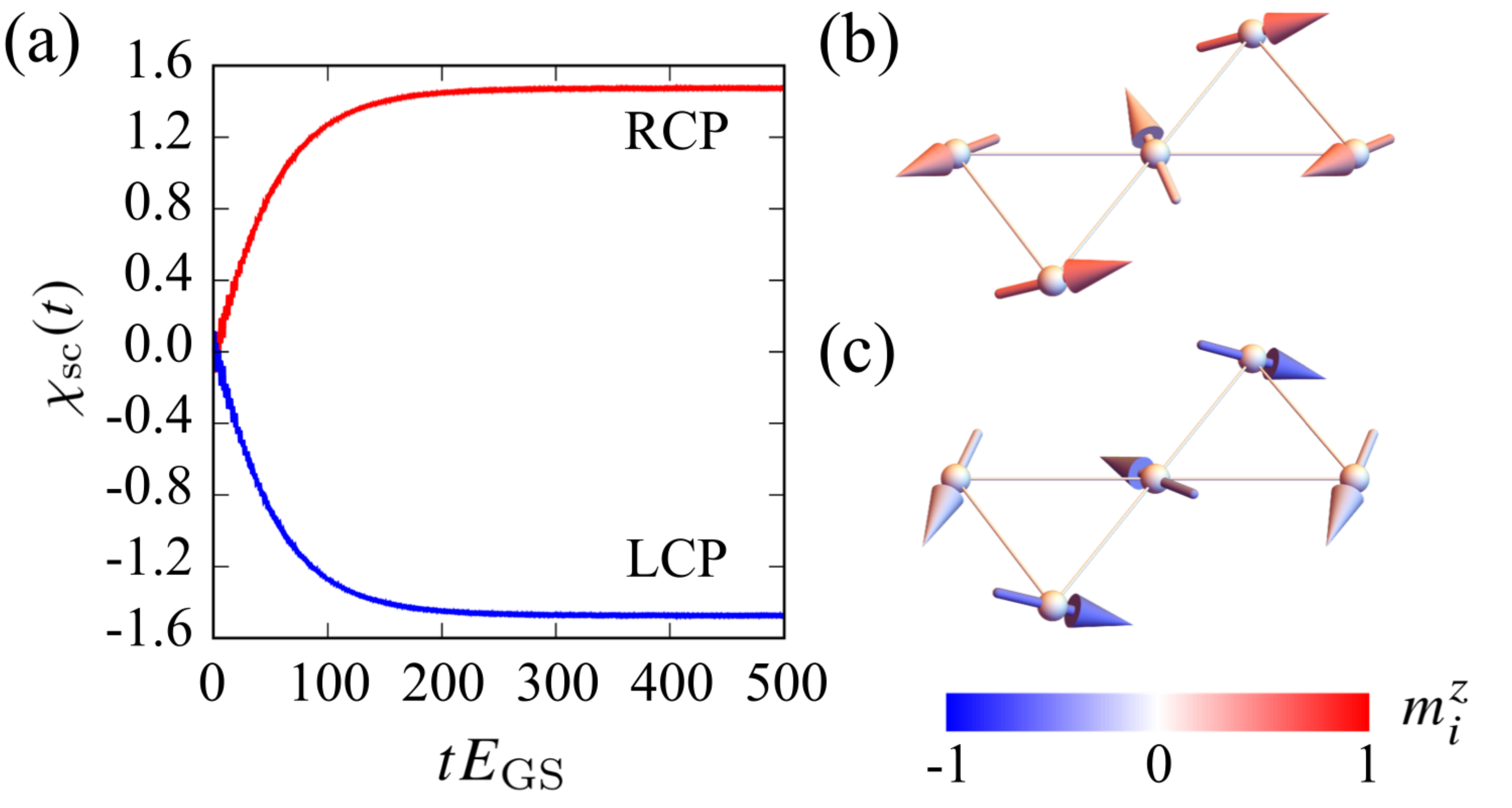} 
\caption{\label{fig:vortex}
(a) Time evolution of the scalar spin chirality in Eq.~(\ref{eq:sc_t}) for the dynamical model in Eq.~(\ref{eq:dynamical_hamiltonian}) with $F^z=-1$, $F^\perp=-0.5$, $D^z=-2$, $E_{\rm d}=0.5E_{\rm GS}$, and $\Omega=4E_{\rm GS}$.
The initial state is set as the vortex state in Eq.~(\ref{eq:conf_vortex}) and the static energy scale is given by $E_{\rm GS}=3|-F^\perp+\sqrt{3}D^z|$.
The red (blue) line shows the time evolution by the electric field with the RCP (LCP).
(b) [(c)]  Snapshot of the magnetic configuration in the NESS by the RCP (LCP).
The arrows and their colors show the local magnetic moments and the $z$ components, respectively.
}
\end{center}
\end{figure}

First, we show the results when we radiate the circularly polarized electric field on the vortex state.
We set $F^z=-1$, $F^\perp=-0.5$, and $D^z=-2$ in the static model.
Then, the ground state becomes the vortex state in Eq.~(\ref{eq:conf_vortex}) and the energy scale is given by $E_{\rm GS}=3|-F^\perp+\sqrt{3}D^z|$.

The red line in Fig.~\ref{fig:vortex}(a) shows a time evolution of the scalar spin chirality in Eq.~(\ref{eq:sc_t}) under the electric field with the RCP, where the field parameters are taken as $E_{\rm d}=0.5E_{\rm GS}$ and $\Omega=4E_{\rm GS}$. 
At $t=0$, the scalar spin chirality is zero because of the coplanar structure in the vortex state.
After the electric field radiation,  the scalar spin chirality is continuously induced and the system reaches a NESS with a positive scalar spin chirality at $t\sim 500E_{\rm GS}^{-1}$.
We show a snapshot of the NESS in Fig.~\ref{fig:vortex}(b), where positive $z$ components of the magnetic moment are induced in the coplanar vortex configuration.
Thus, this magnetic configuration is characterized by the scalar chiral configuration in Eq.~(\ref{eq:conf_L1}).
It is noted that the NESS exhibits a small in-plane magnetization compared to the out-of-plane one by around $10^{-3}$.

Meanwhile, when the polarization is reversed so as to have the LCP, the negative scalar spin chirality is developed, as shown by the blue line in Fig.~\ref{fig:vortex}(a).
After a long time, the system reaches a NESS with a negative scalar spin chirality, whose snapshot is shown in Fig.~\ref{fig:vortex}(c); the NESS corresponds to the scalar chiral state with negative $z$ components of the magnetic moment. 
Similar to the case with the RCP, a small in-plane magnetization also appears in this case.

The microscopic origin of the scalar spin chirality is attributed to the effective field-induced three-spin interactions appearing in the Floquet Hamiltonian in Eq.~(\ref{eq:floquet_hamiltonian}); the negative $\mathcal{T}$ is induced by the electric field with the RCP and it favors the positive scalar spin chirality. 
Indeed, we find that the behavior of the scalar spin chirality in the dynamical model in Eq.~(\ref{eq:dynamical_hamiltonian}) is well fitted by the Floquet model in Eq.~(\ref{eq:floquet_hamiltonian}), as detailed below. 
Similarly, the opposite sign of the scalar spin chirality by the LCP is owing to the positive $\mathcal{T}$.
These results indicate that the sign of the scalar spin chirality is controlled by the polarization of the electric field.

Let us comment on the electric field radiation on the antivortex state in Eq.~(\ref{eq:conf_vortex}), which is stabilized by taking the positive DM interaction $D^z=2$; the other parameters are the same as the previous ones.
In this situation, a NESS with a positive (negative) scalar spin chirality is induced by radiating the electric field with the RCP (LCP), which is similar to the radiation on the vortex state in Fig.~\ref{fig:vortex}(a).
Meanwhile, magnetic configurations of the NESSs are different from those in Figs.~\ref{fig:vortex}(b) and \ref{fig:vortex}(c); with the RCP (LCP) negative (positive) $z$ components of the magnetic moment are induced in the coplanar antivortex configuration.

\begin{figure}[t!]
\begin{center}
\includegraphics[width=1.0\hsize]{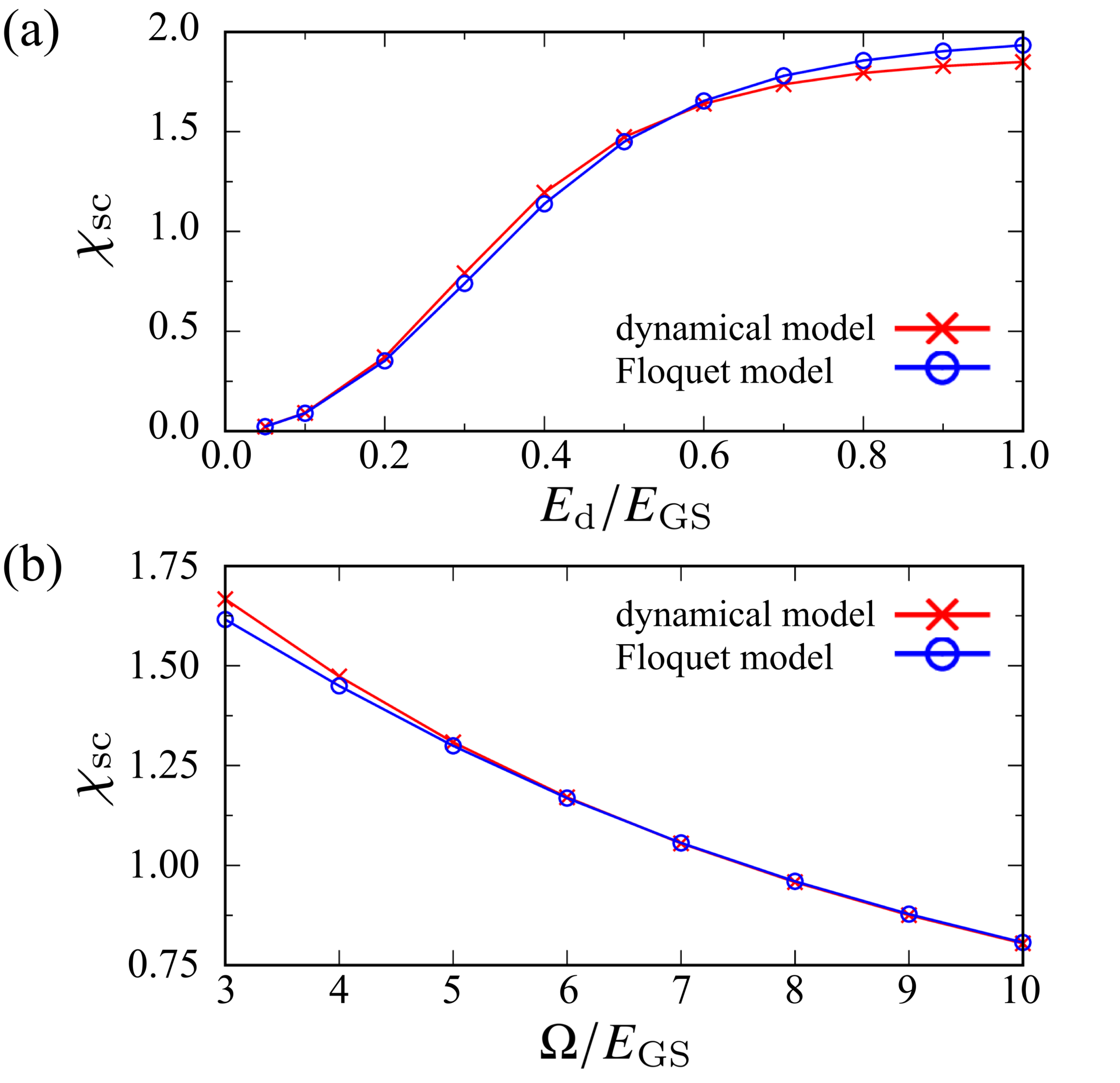} 
\caption{\label{fig:vortex_field}
Electric-field parameter dependence of the averaged scalar spin chirality $\chi_{\rm sc}$ in Eq.~(\ref{eq:sc}) in the dynamical and Floquet models with $F^z=-1$, $F^\perp=-0.5$, $D^z=-2$, and $\delta=1$: (a)  $E_{\rm d}$ dependence at $\Omega=4E_{\rm GS}$ and (b) $\Omega$ dependence at $E_{\rm d}=0.5E_{\rm GS}$.    
The initial state is set as the vortex state in Eq.~(\ref{eq:conf_vortex}) and the static energy scale is given by $E_{\rm GS}=3|-F^\perp+\sqrt{3}D^z|$.
The red (blue) line shows the results for the dynamical model in Eq.~(\ref{eq:dynamical_hamiltonian}) [Floquet model in Eq.~(\ref{eq:floquet_hamiltonian})].
}
\end{center}
\end{figure}

Next, we show the behavior of the scalar spin chirality while changing the amplitude $E_{\rm d}$ and the frequency $\Omega$.
In Fig.~\ref{fig:vortex_field}(a), we show the $E_{\rm d}$ dependence of the averaged scalar spin chirality under the electric field with the RCP and $\Omega=4E_{\rm GS}$; the red and blue lines show the results for the dynamical and Floquet models, respectively.
The results for both models show their good agreement except for the large $E_{\rm d}$ region.
This indicates that the effective field-induced three-spin interactions $\mathcal{T}$ play an important role in inducing the scalar spin chirality $\chi_{\rm sc}$, and higher-order contributions, such as terms proportional to $E_{\rm d}^2\Omega^{-2}$, are almost negligible in the small $E_{\rm d}$ region.
The scalar spin chirality increases with increasing $E_{\rm d}$ proportional to $E_{\rm d}^2$ in the small $E_{\rm d}$ region, which is also consistent with the expression of $\mathcal{T}$ in Eqs.~(\ref{eq:T1})--(\ref{eq:T3}).
When $E_{\rm d}$ is further increased, the scalar spin chirality approaches the maximum value of $\chi_{\rm sc}=2$.

In Fig.~\ref{fig:vortex_field}(b), we show the $\Omega$ dependence of the averaged scalar spin chirality 
under the electric field with the RCP and $E_{\rm d}=0.5E_{\rm GS}$.
Similar to the result in Fig.~\ref{fig:vortex_field}(a), one finds that the results for the dynamical and Floquet models are consistent with each other; the small derivation in the low $\Omega$ region is due to the higher-order contributions in the high-frequency expansion, such as terms proportional to $E_{\rm d}^2\Omega^{-2}$.
The behavior of the scalar spin chirality proportional to $\Omega^{-1}$ is understood from the effective interaction $\mathcal{T}$ in Eqs.~(\ref{eq:T1})--(\ref{eq:T3}).
It is noted that a similar behavior of the scalar spin chirality in Fig.~\ref{fig:vortex_field} is obtained when the electric field radiation on the antivortex state is considered.

\subsection{Radiation on a $z$-FM state}
\label{sec:simulation_FM}

\begin{figure}[t!]
\begin{center}
\includegraphics[width=1.0\hsize]{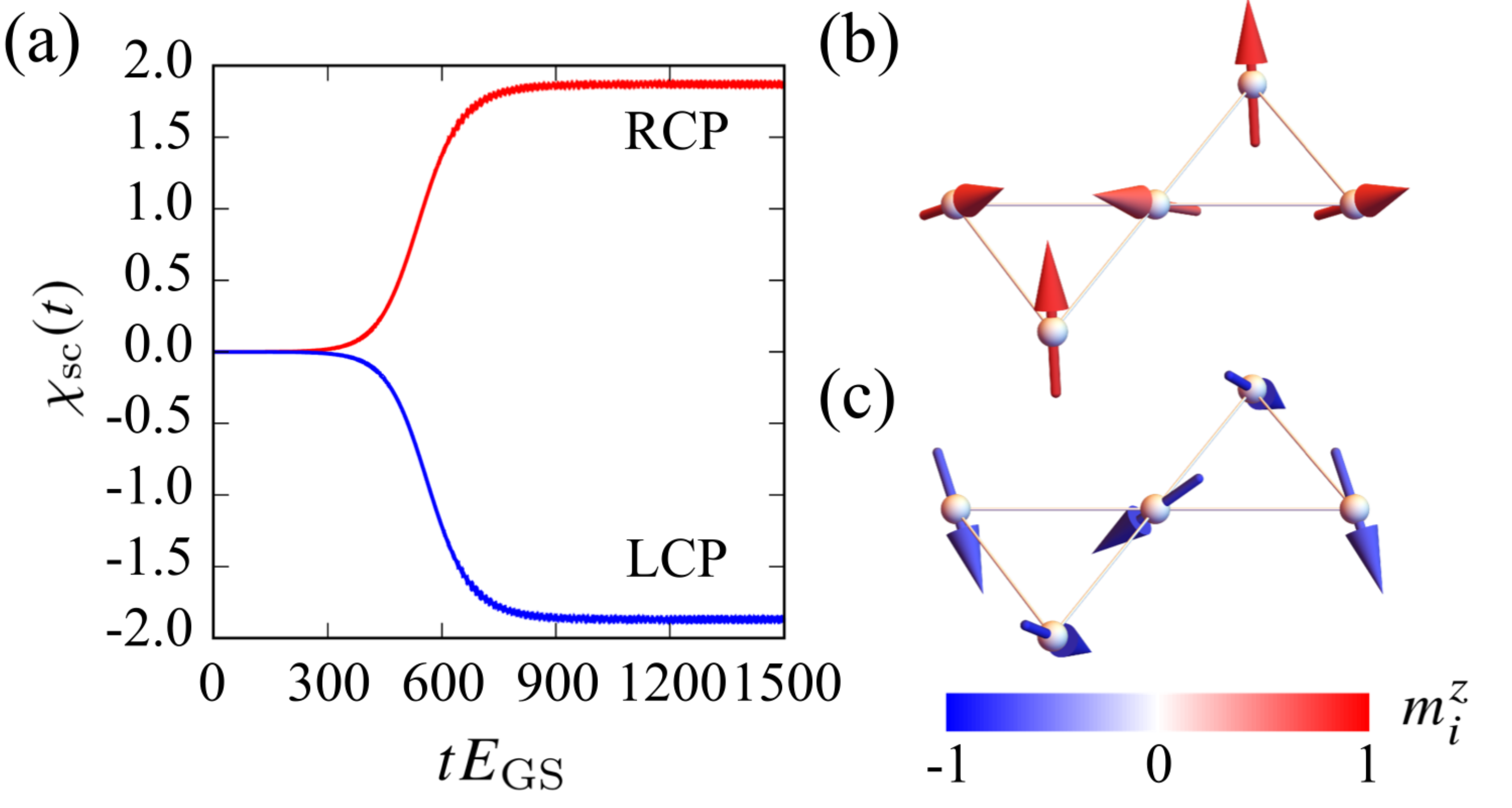} 
\caption{\label{fig:FM}
(a) Time evolution of the scalar spin chirality in Eq.~(\ref{eq:sc_t}) for the dynamical model in Eq.~(\ref{eq:dynamical_hamiltonian}) with $F^z=-1$, $F^\perp=-0.5$, $D^z=-1.2$, $E_{\rm d}=0.5E_{\rm GS}$, and $\Omega=4E_{\rm GS}$.
The initial state is set as the $z$-FM state and the energy scale is given by $E_{\rm GS}=|
6F^z|$.
The red (blue) line shows the time evolution under the electric field with the RCP (LCP).
(b) [(c)] Snapshot of the magnetic configuration in the NESS by the RCP (LCP).
The arrows and their colors show the local magnetic moments and the $z$ components, respectively.
}
\end{center}
\end{figure}

Next, we investigate the effect of the circularly polarized electric field radiation on the $z$-FM state.
We set $F^z=-1$, $F^\perp=-0.5$, and $D^z=-1.2$ in the static model, where 
the ground state becomes the $z$-FM state in Eq.~(\ref{eq:conf_z}) and the energy scale is given by $E_{\rm GS}=|6F^z|=6$.

We show a time evolution of the scalar spin chirality after introducing the electric field with the RCP (LCP), $E_{\rm d}=0.5E_{\rm GS}$, and $\Omega=4E_{\rm GS}$ by the red (blue) line in Fig.~\ref{fig:FM}(a).
The system irradiated by the electric field with the RCP (LCP) reaches a NESS with a positive (negative) scalar spin chirality at $t\sim 1500E_{\rm GS}$.
The sign of the scalar spin chirality is explained by the sign of the field-induced three-spin interactions $\mathcal{T}$.
This indicates that the scalar spin chirality is controlled by the polarization, which is similar to the result for the vortex (antivortex) state in Sec.~\ref{sec:simulation_vortex}.
We show a snapshot of the NESS induced by the electric field with the RCP (LCP) in Fig.~\ref{fig:FM}(b) [Fig.~\ref{fig:FM}(c)], where a coplanar vortex configuration is additionally induced in the $z$-FM state with the positive (negative) $z$ components so as to have the positive (negative) scalar spin chirality.
These magnetic configurations correspond to the scalar chiral configuration, which is realized as the static ground state at the phase boundary L1 in Fig.~\ref{fig:GS}(a).

Let us comment on the initial states for Fig.~\ref{fig:FM}. 
We introduce the small random deviation from the magnetic configuration in Eq.~(\ref{eq:conf_z}) at $t=0$ as follows:
We set the initial state as $\boldsymbol{m}_j=[\sqrt{1-(m_j^z)^2}\cos\phi_j,\sqrt{1-(m_j^z)^2}\sin\phi_j,\pm(1-a_j)]$, where $a_j$ ($0\le a_j \le10^{-3}$) and $\phi_j$ ($0< \phi_j \le 2\pi$) are random variables~\footnote{We confirm that similar results are obtained for smaller fluctuations. e.g., $0\le a_j \le10^{-4}$ and $0\le a_j \le10^{-5}$.}. 
This is because no effective magnetic field is generated under the electric field in Eq.~(\ref{eq:Beff}) when the magnetic configuration is initially set as $\boldsymbol{m}_j=(0,0,\pm 1)$.
It is also noted that the positive (negative) $z$ component of the magnetic moment in the initial state is related to the finite scalar spin chirality by the electric field with the RCP (LCP). 
In other words, no scalar spin chirality is induced when the electric field with the RCP (LCP) is applied to the $z$-FM state with negative (positive) $z$-spin polarization.

\begin{figure}[t!]
\begin{center}
\includegraphics[width=1.0\hsize]{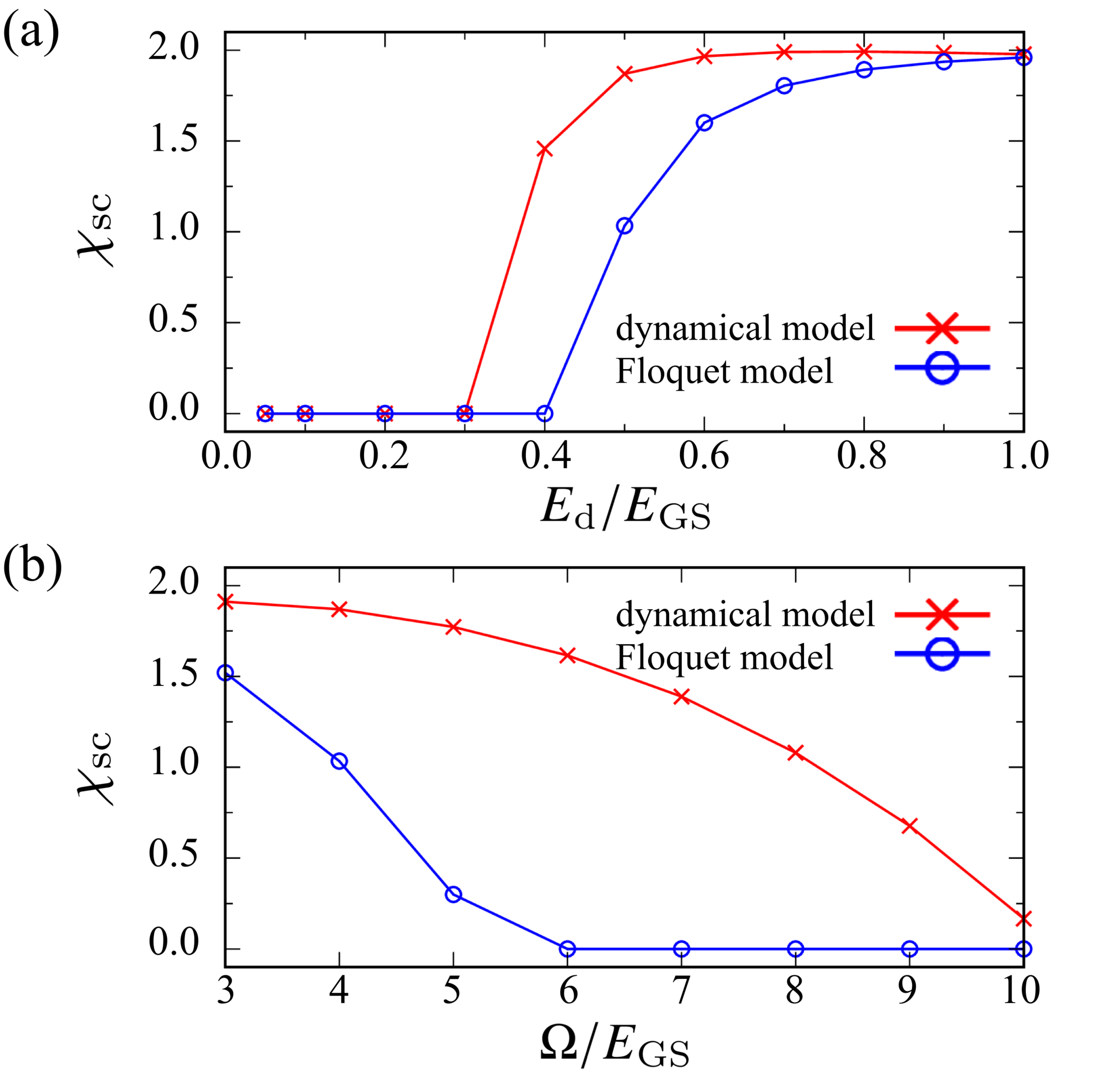} 
\caption{\label{fig:FM_field}
Electric-field parameter dependence of the averaged scalar spin chirality $\chi_{\rm sc}$ in Eq.~(\ref{eq:sc})  in the dynamical and Floquet models with $F^z=-1$, $F^\perp=-0.5$, $D^z=-1.2$, and $\delta=1$: (a)  $E_{\rm d}$ dependence at $\Omega=4E_{\rm GS}$ and (b) $\Omega$ dependence at $E_{\rm d}=0.5E_{\rm GS}$.    
The initial state is set as the $z$-FM state and the static energy scale is given by $E_{\rm GS}=|6F^z|$.
The red (blue) line shows the results for the dynamical model in Eq.~(\ref{eq:dynamical_hamiltonian}) [Floquet model in Eq.~(\ref{eq:floquet_hamiltonian})].
}
\end{center}
\end{figure}

In Fig.~\ref{fig:FM_field}(a), we show the $E_{\rm d}$ dependence of the averaged scalar spin chirality induced by the electric field with the RCP and $\Omega=4E_{\rm GS}$; the red and blue lines show the results for the dynamical and Floquet models, respectively. 
In the dynamical (Floquet) model, the scalar spin chirality is not induced in the $E_{\rm d}\leq 0.3$ ($E_{\rm d}\leq 0.4$) region.  
As $E_{\rm d}$ increases, both models exhibit nonzero scalar spin chirality.
The enhancement of the scalar spin chirality for large $E_{\rm d}$ is owing to the relatively large three-spin interactions $\mathcal{T}\propto E_{\rm d}^2$.  
This qualitative behavior is similar to that in Fig.~\ref{fig:vortex_field}(a), while there is no quantitative agreement between the dynamical and Floquet models.
The scalar spin chirality in the dynamical model tends to be larger than that in the Floquet model, which might be attributed to the fact that the higher-order contribution than $E^2_{\rm d}\Omega^{-1}$ included in the dynamical model favors the scalar spin chirality.

We also show the $\Omega$ dependence of the averaged scalar spin chirality induced by the electric field with the RCP and $E_{\rm d}=0.5E_{\rm GS}$ in Fig.~\ref{fig:FM_field}(b).
At $\Omega=3$, the scalar spin chirality is induced in the dynamical and Floquet models.
The scalar spin chirality decreases monotonically with increasing $\Omega$ due to the three-spin interactions $\mathcal{T}\propto \Omega^{-1}$.
By further increasing $\Omega$, the scalar spin chirality approaches zero.  
Similar to the $E_{\rm d}$ dependence, the results for the dynamical and Floquet models show qualitative agreement, but not quantitative agreement.

\subsection{Static-parameter dependence}
\label{sec:simulation_parameter}

\begin{figure}[t!]
\begin{center}
\includegraphics[width=1.0\hsize]{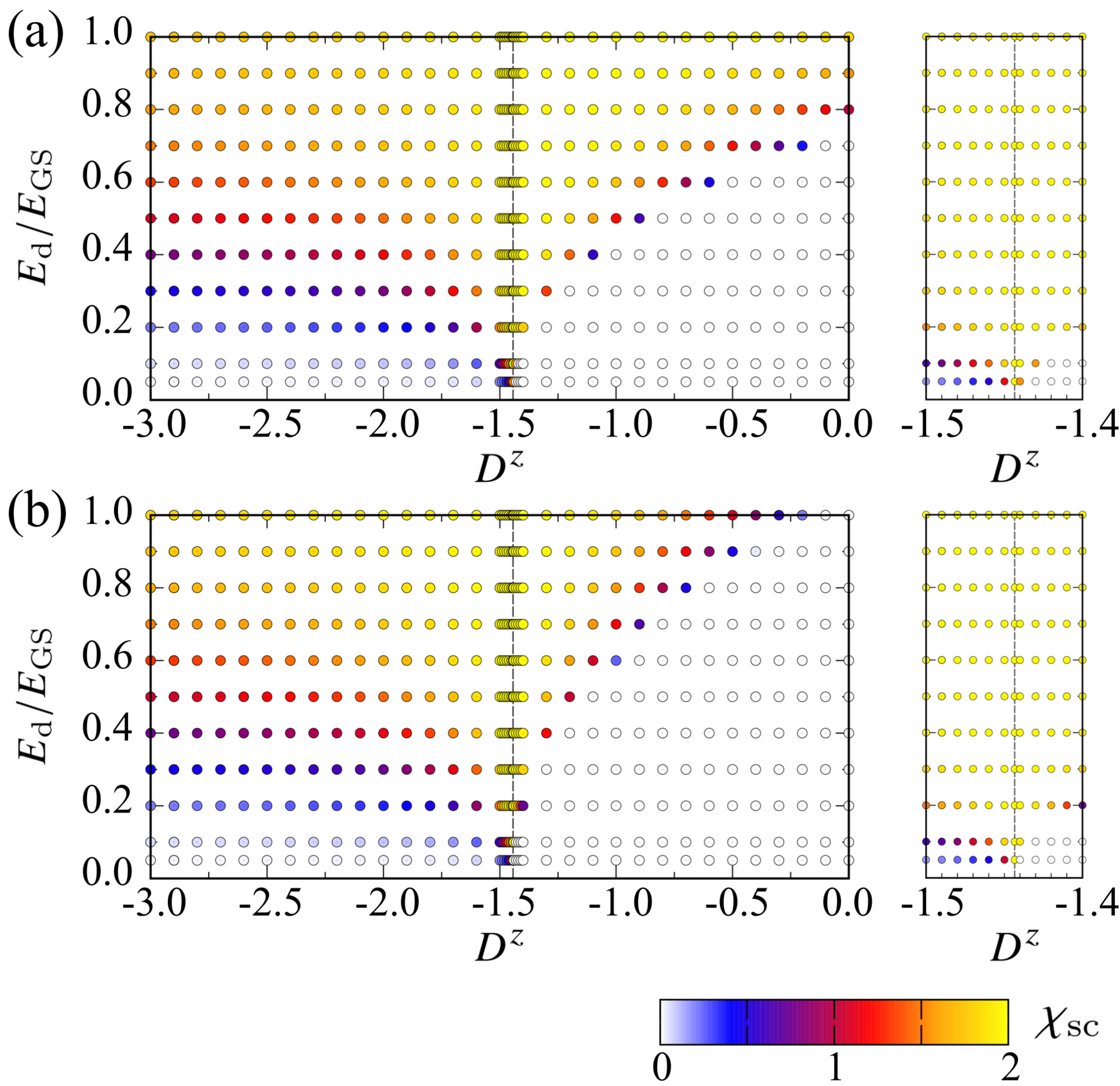} 
\caption{\label{fig:PD}
Parameter dependence of the averaged scalar spin chirality $\chi_{\rm sc}$ in Eq.~(\ref{eq:sc}) in (a) the dynamical model in Eq.~(\ref{eq:dynamical_hamiltonian}) and (b) the Floquet model in Eq.~(\ref{eq:floquet_hamiltonian}).
We change $D^z$ and $E_{\rm d}$ while fixing $F^z=-1$, $F^\perp=-0.5$, $\delta=1$, and $\Omega=4E_{\rm GS}$.
The initial states are set as the vortex state for $D^z<D^{z*}=-2.5/\sqrt{3}$, the scalar chiral state for $D^z=D^{z*}$, and the $z$-FM state for $D^z>D^{z*}$.
The vertical dashed line shows $D^z=D^{z*}$. 
The right panel shows the results around $D^{z*}$. 
}
\end{center}
\end{figure}

Finally, we discuss the averaged scalar spin chirality induced by the circularly polarized electric field in terms of $D^z$ and $E_{\rm d}$.
In Fig.~\ref{fig:PD}(a) [\ref{fig:PD}(b)], we show the averaged scalar spin chirality with changing $D^z$ and $E_{\rm d}$ in the dynamical (Floquet) model with $F^z=-1$, $F^\perp=-0.5$, $\delta=1$, and $\Omega=4E_{\rm GS}$~\footnote{We confirm that a similar result is obtained in the models with $(F^z,F^\perp)=(-1,0)$ and $(-1,0.5)$. A similar result is also obtained in the model with the LCP ($\delta=-1$)}.
The initial states are set as the vortex state for $D^z<D^{z*}=-2.5/\sqrt{3}$, the scalar chiral state in Eq.~(\ref{eq:conf_L1}) with $\chi_{\rm sc}=2$ for $D^z=D^{z*}$, and the $z$-FM state for $D^z>D^{z*}$.
The results for both models show the similar parameter dependence.

For $D^z<D^{z*}$, the radiation of the circularly polarized electric field on the vortex state induces the scalar spin chirality irrespective of $D^z$ and $E_{\rm d}$.
The induced scalar spin chirality becomes large in the large $E_{\rm d}$ region for a fixed $D^z<D^{z*}$, as discussed in Sec.~\ref{sec:simulation_vortex}. 
In addition, one finds that the scalar spin chirality tends to become large as $D^z$ increases. 
Especially, as shown in the right panels of Figs.~\ref{fig:PD}(a) and \ref{fig:PD}(b), the scalar spin chirality is strongly enhanced in the vicinity of $D^{z*}$.
This might be attributed to the fact that the magnetic configuration with the large scalar spin chirality is realized when the weights of the vortex and $z$-FM states in the scalar chiral configuration in Eq.~(\ref{eq:conf_L1}) are comparable to each other.

In the region for $D^z>D^{z*}$, the $z$-FM state is stabilized as the ground state within the static model.
The scalar spin chirality is also induced when the amplitude of the electric field is larger than the critical value of $E_{\rm d}^*$, as discussed in Sec.~\ref{sec:simulation_FM}.
Such a tendency is found for any $D^z$.  
For a fixed $E_{\rm d}$, the scalar spin chirality is enhanced near $D^{z*}$ for the same reason in the $D^z<D^{z*}$ region; see the right panels of Figs.~\ref{fig:PD}(a) and \ref{fig:PD}(b).

At $D^z=D^{z*}$, the static ground state is a superposition of the vortex and $z$-FM states, where magnetic states with $|\chi_{\rm sc}|\le 2$ are energetically degenerate.  
By radiating the electric field in such a region, this degeneracy is lifted by the three-spin interactions $\mathcal{T}$; the magnetic state with $\chi_{\rm sc}= 2$ has the lowest energy in the Floquet model.
Such a tendency holds for any $E_{\rm d}$.

\section{Summary and Discussion}
\label{sec:summary}

In summary, we have studied how to generate and control the scalar spin chirality by the circularly polarized electric field by exemplifying the classical kagome magnet.
By taking into account the coupling of the electric field and the spin-dependent electric polarization, we have elucidated the generation of the scalar spin chirality irrespective of the collinear and coplanar spin configurations in the ground state.
We have shown that the microscopic origin of the scalar spin chirality is the effective field-induced three-spin interactions, which can be analytically obtained in the high-frequency regime based on the Floquet formalism.
Furthermore, we have shown that the sign and magnitude of the scalar spin chirality are controlled by the amplitude, frequency, and polarization of the circularly polarized electric field. 
We have also shown that the scalar spin chirality tends to be enhanced near the phase boundary between the coplanar vortex phase and the collinear $z$-FM phase.
Our results indicate the possibility of inducing the scalar spin chirality by the electric field rather than the magnetic field, which would provide an alternative root of controlling topological spin textures.

Let us discuss a possible experimental situation.
In order to induce the scalar spin chirality by the circularly polarized electric field, considering insulating magnets is important since the coupling between the electric field and the electric polarization plays an essential role.
Assuming a magnet with an exchange interaction of 1~meV, our Floquet analysis in the high-frequency regime is valid for $\Omega > 1$ THz; a typical magnitude of the terahertz electric field is estimated as $E_0=1$--$10$ MV/cm.  
Meanwhile, $\Omega$ should be smaller than a band gap in the order of eV to avoid heating effects by the electric field radiation.

\begin{acknowledgments}
We thank K. Shimizu, S. Okumura, Y. Kato, and Y. Motome for fruitful discussions.
This research was supported by JSPS KAKENHI Grants Numbers JP19K03752, JP19H01834, JP21H01037, JP22H04468, JP22H00101, JP22H01183, JP23KJ0557, JP23H04869, JP23K03288, and by JST PRESTO (JPMJPR20L8) and JST CREST (JPMJCR23O4). 
R.Y. was supported by Forefront Physics and Mathematics Program to Drive Transformation (FoPM) and JSPS Research Fellowship.
\end{acknowledgments}

\bibliography{main.bbl}
\end{document}